\documentclass[12pt]{iopart}
\usepackage{subfigure,psfrag}
\usepackage[dvipdfm]{graphicx}
\usepackage{color}
\usepackage{ulem}

\usepackage{epstopdf} 

\def\0{0}
\def\1{1}

\def\U#1{{\rm #1}}
\def\u#1{_{\rm #1}}

\newcommand{\bra}[1]{\langle #1 |}
\newcommand{\ket}[1]{| #1 \rangle}
\newcommand{\braeq}[1]{\left\langle #1 \right|}
\newcommand{\keteq}[1]{\left| #1 \right\rangle}
\newcommand{\ketbra}[2]{| #1 \rangle \langle #2 |}

\def\tr{\U{tr}}

\def\subnumcases{\let\numc@setsub\subequations 
   \let\numc@resetsub\endsubequations  \numcases}

\begin{document}

\title[Universal gates for transforming multipartite entangled Dicke states]{Universal gates for transforming multipartite entangled Dicke states}

\author{Toshiki Kobayashi$^{1,*}$, Rikizo Ikuta$^1$, \c{S}ahin Kaya \"{O}zdemir$^{2,\dag}$, \\ Mark Tame$^{3,\ddag}$, Takashi Yamamoto$^1$, Masato Koashi$^4$, \\ and Nobuyuki Imoto$^1$}
\address{$^1$Graduate School of Engineering Science, Osaka University,
Toyonaka, Osaka 560-8531, Japan}
\address{$^2$Dept. of Electrical and Systems Engineering,
Washington University, St. Louis, MO 63130, USA}
\address{$^3$School of Chemistry and Physics, University of KwaZulu-Natal, Durban 4001, South~Africa}
\address{$^4$Photon Science Center,
The University of Tokyo, Bunkyo-ku, 113-8656, Japan}

\ead{$^*$kobayashi-t@qi.mp.es.osaka-u.ac.jp, \\ $^\dag$ozdemir@ese.wustl.edu, $^\ddag$markstame@gmail.com}
\begin{abstract}
We determine the minimal number of qubits 
that it is necessary to have access to 
in order to transform Dicke states into other Dicke states. 
In general, the number of qubits in Dicke states cannot 
be increased via transformation gates 
by accessing only a single qubit, 
in direct contrast to other multipartite entangled states
such as GHZ, W and cluster states.
We construct a universal optimal gate 
which adds spin-up qubits or spin-down qubits to any Dicke state 
by minimal access. 
We also show the existence of a universal gate 
which transforms any size of Dicke state 
as long as 
it has access to at least the required number of qubits. Our results have important consequences for the generation of Dicke states in physical systems such as ion traps, all-optical setups and cavity-QED settings where they can be used for a variety of quantum information processing tasks.
\end{abstract}
\maketitle
\section{Introduction}
Entanglement is a key resource facilitating a wide range of emerging
quantum technologies, such as quantum computing~\cite{Ladd}, communication~\cite{Kimble,cloning} and sensing~\cite{metrology}. It has been well established theoretically~\cite{qip} and
experimentally demonstrated between various particles, including photons, atoms and ions~\cite{Monroe}.
Entanglement between two particles~\cite{EPR} has been routinely prepared and used in different physical systems for a variety of tasks~\cite{Ladd,Kimble,cloning,metrology,qip}.
However, in order to make full use of the power of entanglement
for quantum technologies and to probe deeper into the foundations of quantum mechanics,
there has been an increasing push toward making larger numbers
of particles entangled with each other.
As the number of particles increases beyond two, different types of entangled states
that cannot be converted into each other using local operations
and classical communication~(LOCC)~\cite{Dur} emerge. Greenberger-Horne-Zeilinger~(GHZ)~\cite{GHZ},
cluster~\cite{cluster}, Dicke~\cite{Dicke} and W states~\cite{W},
are examples of such inequivalent classes. Dicke states in particular provide a rich variety of structurally complex states
among many particles and hold great promise
for a wide range of applications in quantum information. Recent experiments have demonstrated the generation of these states in physical systems such as ion traps~\cite{Iontrap,Iontrap2}, all-optical setups~\cite{WExp,Wex,WExp3,DickeExp1,DickeExp2,DickeExp3,DickeExp4,DickeExp5} and cavity quantum electrodynamic settings~\cite{WExp2}. Despite these impressive demonstrations, the complexity of Dicke states makes their preparation and manipulation difficult.
Thus, understanding the limits for preparing and manipulating large multipartite entangled versions are of great interest and urgently needed.

In this paper, 
we derive the minimal number of qubits
that it is necessary to have access to 
in order to expand and reduce any given Dicke state. 
We show that, unlike W~\cite{Wex}, 
GHZ and cluster states~\cite{Browne}, 
Dicke states in general cannot be transformed by local access
to only a single qubit. 
We consider gates for transforming Dicke states by minimal access. 
In the case of the expansion of W states, by accessing only one qubit 
there is a universal optimal gate 
which can expand any size of W state with maximum success probability. 
Similarly to this case, 
we derive a universal optimal gate 
which adds either spin-up or spin-down qubits 
to Dicke states by minimal access. 
We then construct universal gates 
which can add or subtract given numbers of spin-up and spin-down qubits 
with a nonzero success probability, 
regardless of the size of an initial Dicke state. 
Our work has important implications 
for assessing the amount of control one needs 
in the preparation and manipulation of Dicke states in physical systems
for a range of quantum information applications, 
such as quantum algorithms~\cite{Grov}, 
quantum games~\cite{game}, testing efficient tomographic techniques~\cite{DickeExp5} and 
multi-agent quantum networking~\cite{DickeExp1,DickeExp2,DickeExp3,DickeExp4}.
\section{Necessary condition for transforming a Dicke state}
An $N$-qubit Dicke state with $M_1$ excitations is the equally weighted
superposition of all permutations of $N$-qubit product states
with $M_1$ spin-up~($\ket{1}$)
and $M_0=N-M_1$ spin-down~($\ket{0}$),
and is written as
\begin{equation}
\keteq{D_N^{M_1}}
=(C_N^{M_1})^{1/2}\hat{P}\keteq{M_0,M_1},
\label{eq:dicke1}
\end{equation}
where
$\ket{M_0,M_1}\equiv \ket{M_0}\ket{M_1}$ with
$\ket{M_i}\equiv \ket{i}^{\bigotimes M_i}$ for $i=\{ 0,1\}$, 
$C_N^{M_1}\equiv N!/(M_1!(N-M_1)!)$, 
and $\hat{P}$ is a projector onto the symmetric subspace
with respect to the permutation of any two particles. 
For example, 
$\hat{P}\ket{2,0}=\ket{00}=\ket{D_2^0}$, 
$\hat{P}\ket{1,1}=(\ket{01}+\ket{10})/2=\ket{D_2^1}/\sqrt{2}$, 
and $\hat{P}\ket{0,2}=\ket{11}=\ket{D_2^2}$. 
Eq.~(\ref{eq:dicke1}) describes general symmetric Dicke states and the theory we develop covers this entire class.
\begin{figure}[t]
\vskip0.5cm
 \begin{center}
  \scalebox{0.6}{\includegraphics{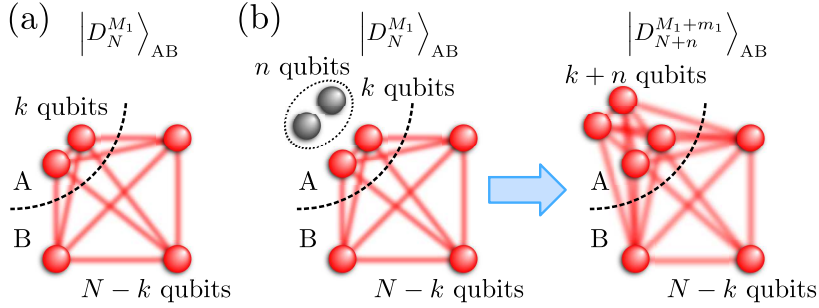}}
 \end{center}
 \caption{(a)
 The Dicke state $\ket{D_N^{M_1}}\u{AB}$
 shared between physical subsystems A and B, which hold $k$ and
$N-k$ qubits, respectively. The qubits represent ions, photons or atoms depending on the physical setting~\cite{Iontrap,Iontrap2,WExp,Wex,WExp3,DickeExp1,DickeExp2,DickeExp3,DickeExp4,DickeExp5,WExp2}. (b) Expansion of $\ket{D_N^{M_1}}\u{AB}$ to $\ket{D_{N+n}^{M_1+m_1}}\u{AB}$
 by accessing only $k$ qubits in A. In the case of reduction,
 $n$ qubits are deleted from  A.
 }
 \label{fig:img}
\end{figure}

We assume that $\ket{D_N^{M_1}}$ is shared
between two physical subsystems A and B, and denote this as
$\ket{D_N^{M_1}}\u{AB}$, with subsystem A
holding a total of $k$ qubits
and subsystem B holding the remaining qubits, as shown
in Fig.~\ref{fig:img}~(a). 
Here we derive the minimum number $k$ of qubits that it is necessary to 
have access to in order to transform the state $\ket{D_N^{M_1}}\u{AB}$
into a state $\ket{D_{N+n}^{M_1+m_1}}\u{AB}$,
where $|n|$ is the total number of
qubits added for $n > 0$ and
deleted for $n < 0$, and similarly
$|m_1|$ is the added or deleted number of qubits in $\ket{1}$,
while $m_0\equiv n-m_1$ represents the added or deleted number of qubits 
in $\ket{0}$. For the trivial cases of 
$M_0=0~(M_0+m_0=0)$ and $M_1=0~(M_1+m_1=0)$, 
the states of system AB are product states 
$\ket{1\ldots 1}$ and $\ket{0\ldots 0}$, respectively. 
In the following, 
we will study only the nontrivial cases where $M_0 > 0$, $M_1 > 0$, 
$M_0+m_0 > 0$ and $M_1+m_1 > 0$. 
We consider a local transformation scenario 
in which access to subsystem B is forbidden, 
and the transformation task is carried out 
by collectively manipulating the $k$ qubits 
of subsystem A only~[see Fig.~\ref{fig:img}~(b)]. This limited-access scenario allows us to investigate the requirements for the number of qubits that one would need control over in a given physical system. In this scenario,
the whole system after the transformation
is composed of $N-k$ qubits in subsystem B
and $k+n$ qubits in subsystem A.
Thus, we have $N+n\geq N-k$, namely 
$N\geq k\geq -n$ is necessary. In other words, the new total number of qubits must be at least as big as the number of qubits in subsystem B to which access is forbidden.

We now derive a necessary condition for the transformation of Dicke states. 
When we consider the superposition of pure product states
for the Dicke state in the computational basis, 
the minimum number of spin-up qubits in subsystem B is obtained by maximizing the number of spin-up qubits in subsystem A, and is 
given by $\alpha=\max\{M_1-k,0\}$ 
for the initial Dicke state $\ket{D_N^{M_1}}$ 
and $\alpha'=\max\{M_1+m_1-k-n,0\}= \max\{M_1-k-m_0,0\}$ 
for the final Dicke state $\ket{D_{N+n}^{M_1+m_1}}$. 
Since subsystem B is left untouched in the transformation, the minimum number of spin-up qubits cannot decrease in subsystem B.
However it may increase, for example if qubits are deleted from subsystem A. Thus, the relation 
$\alpha'\geq \alpha$ should hold.
This means that $k \geq M_1$ is a necessary condition for the transformation with $m_0 > 0$. Since a similar argument holds for the transformation with $m_1 > 0$, 
we have
\begin{eqnarray}
  \label{eq:kcondition}
  k\geq\left\{ \begin{array}{ll}
    M_1 & {\rm for}~~ m_0 > 0, \\
    M_0 & {\rm for}~~ m_1 > 0,
  \end{array} \right.
\end{eqnarray}
as a necessary condition 
for transforming a Dicke state to another Dicke state. In other words, for spin-down (spin-up) qubits to be added, the necessary condition is that the number of qubits in subsystem A must be at least as big as the number of spin-up (spin-down) qubits in the total system.
Note that for other cases, we have 
\begin{eqnarray}
k\geq -n\ \ {\rm for}~~ m_0 \leq 0\ {\rm and}~~ m_1\leq 0\label{eq:k3}. 
\end{eqnarray}
This is a trivial condition that the number of qubits in subsystem A 
should be at least as big as the total number of qubits being deleted.

\section{Sufficient condition for transforming a Dicke state}
Here we show that conditions~(\ref{eq:kcondition}) and~(\ref{eq:k3})
are sufficient conditions for transformation of a Dicke state 
to another Dicke state for any given physical system, 
and we derive the maximum probability for the transformation. 
We first decompose the Dicke state in Eq.~(\ref{eq:dicke1}) 
by using the symmetric bases in subsystems A and B. 
When we expand $C_N^{M_1}\hat{P}\ket{M_0,M_1}\u{AB}$ 
in the computational basis, 
it is given by the sum of $C_N^{M_1}$ terms with unit amplitude. 
From these terms, 
we select those that have $j$ spin-up qubits in subsystem B. 
The sum of these selected terms is given by 
$C_{k}^{M_1-j}\hat{P}\ket{k-(M_1-j), M_1-j}\u{A}
C_{N-k}^{j}\hat{P}\ket{(N-k)-j,j}\u{B}$. 
Thus we can write 
$C_N^{M_1}\hat{P}\ket{M_0,M_1}\u{AB}
=
\sum_{j=\alpha}^\beta
C_{k}^{M_1-j}
\hat{P}\ket{k-(M_1-j),M_1-j}\u{A}
C_{N-k}^{j}\hat{P}\ket{(N-k)-j,j}\u{B}$, 
where the range of the summation over $j$ is given by 
\begin{equation}
\alpha=\max\{ M_1-k, 0\},\quad 
\beta=\min\{ N-k, M_1\}
\label{eq:alphabeta}
\end{equation}
and $\beta$ is calculated using similar methods to the derivation of $\alpha$. Using this decomposition, we rewrite Eq.~(\ref{eq:dicke1}) as 
\begin{equation}
\keteq{D_N^{M_1}}\u{AB}
=\sum_{j=\alpha}^{\beta}\sqrt{\frac{C_{k}^{M_1-j}C_{N-k}^{j}}{C_N^{M_1}}}
 \keteq{D_k^{M_1-j}}\u{A}\keteq{D_{N-k}^{j}}\u{B}.
\label{eq:dec1}
\end{equation}

In the following, we treat the case with $k>-n$ and the case with $k=-n$
separately. In each case, assuming the conditions (\ref{eq:kcondition}) and (\ref{eq:k3}), we show that the transformation is optimally achievable at a nonzero probability $p\u{max}$.

For $k> -n$,
decomposition of the desired state $\ket{D_{N+n}^{M_1+m_1}}\u{AB}$ 
obtained from the transformation is 
\begin{equation}
\keteq{D_{N+n}^{M_1+m_1}}\u{AB}
=\sum_{j=\alpha'}^{\beta'}
\sqrt{\frac{C_{k+n}^{M_1+m_1-j}C_{N-k}^{j}}{C_{N+n}^{M_1+m_1}}}
 \keteq{D_{k+n}^{M_1+m_1-j}}\u{A}\keteq{D_{N-k}^{j}}\u{B}
\label{eq:dec2}
\end{equation}
with $\alpha'=\max\{ M_1-k-m_0, 0\}$ 
and $\beta'=\min\{ N-k, M_1+m_1\}$. Again, $\beta'$ is calculated similarly to $\alpha'$.
Since access is allowed only to subsystem A, 
the marginal state in subsystem B 
does not change through the transformation process, 
which implies the relation 
$\tr\u{A}(\ketbra{D_N^{M_1}}{D_N^{M_1}})
=p \tr\u{A}(\ketbra{D_{N+n}^{M_1+m_1}}{D_{N+n}^{M_1+m_1}})
+(1-p)\hat{\rho}\u{f}^\U{B}$ must hold.
Here $p$ is the success probability of the transformation and
$\hat{\rho}\u{f}^\U{B}$ is the state of subsystem B
when the transformation fails. 
From Eq.~(\ref{eq:dec1}),
$\tr\u{A}(\ketbra{D_N^{M_1}}{D_N^{M_1}})$ and
$\tr\u{A}(\ketbra{D_{N+n}^{M_1+m_1}}{D_{N+n}^{M_1+m_1}})$
are diagonalized
by
the basis $\{ \ket{D_{N-k}^j}\u{B}\}_{0\leq j\leq N-k}$. 
Thus, from 
the positivity of $\hat{\rho}\u{f}^\U{B}$, 
we have $p\leq p\u{max}$, 
where 
\begin{equation}
p\u{max} \equiv 
q\u{min}
C_{N+n}^{M_1+m_1}/C_N^{M_1}
\label{eq:Pmax}
\end{equation}
with 
\begin{equation}
q\u{min}\equiv 
\min_{\alpha' \leq j\leq \beta'} q_j 
\label{eq:qmin0}
\end{equation}
and 
\begin{equation}
q_j \equiv
C_k^{M_1-j}/C_{k+n}^{M_1+m_1-j}. 
\label{eq:qj}
\end{equation}
Here it should be understood that 
$C_0^0\equiv 1$, and $C_k^{M_1-j}=0$ for $M_1-j < 0$ and $M_1-j > k$. 
Since we are assuming conditions (\ref{eq:kcondition}) and (\ref{eq:k3}), 
we have $\alpha' \geq \alpha$ and $\beta' \leq \beta$, 
resulting in $p\u{max} > 0$. 
Under these conditions, 
we construct a gate $\mathcal{M}\u{A}$ which achieves 
the upper bound on the success probability in Eq.~(\ref{eq:Pmax}). 
The gate $\mathcal{M}\u{A}$ is composed of 
a success operator $\hat{M}\u{s}$ and 
a failure operator $\hat{M}\u{f}$ satisfying 
$\hat{M}\u{s}^\dagger\hat{M}\u{s}
+ \hat{M}\u{f}^\dagger\hat{M}\u{f} = \hat{I}$. 
We define $\hat{M}\u{s}$ by 
\begin{equation}
\hat{M}\u{s}\equiv
\sum_{j=\alpha'}^{\beta'}\sqrt{q\u{min}\frac{C_{k+n}^{M_1+m_1-j}}
{C_k^{M_1-j}}}
\keteq{D_{k+n}^{M_1+m_1-j}}\u{AA}\hspace{-0.1cm}\braeq{D_k^{M_1-j}}. 
\label{eq:Ms}
\end{equation}
From Eqs.~(\ref{eq:qmin0}) and (\ref{eq:qj}), 
no coefficients of $\hat{M}\u{s}^\dagger\hat{M}\u{s}$ are 
larger than $1$, and thus $\mathcal{M}\u{A}$ is 
a valid measurement process for any given physical system. 
From Eqs.~(\ref{eq:dec1})-(\ref{eq:Pmax}) and (\ref{eq:Ms}), 
we have 
$\hat{M}\u{s}\ket{D_{N}^{M_1}}\u{AB}=\sqrt{p\u{max}}\ket{D_{N+n}^{M_1+m_1}}\u{AB}$.
As a result, for $k>-n$, the maximum probability of
the transformation is given by $p\u{max}$ defined in Eq. (\ref{eq:Pmax}),
which is nonzero when conditions (\ref{eq:kcondition}) and (\ref{eq:k3}) hold.

For the case of $k=-n$, condition (\ref{eq:kcondition}) implies $m_0\leq0$ and $m_1\leq0$ because $k\geq M_1 \geq -m_1 > -m_0-m_1=-n$ for $m_0>0$ and $k \geq M_0 \geq -m_0 > -m_0-m_1=-n$ for $m_1 > 0$.
Thus the desired state after the transformation is 
$\ket{D_{N-|n|}^{M_1-|m_1|}}\u{B}$. 
From the relation  
$\tr\u{A}(\ketbra{D_N^{M_1}}{D_N^{M_1}})
=p \ket{D_{N-|n|}^{M_1-|m_1|}}\u{BB}\bra{D_{N-|n|}^{M_1-|m_1|}}
+(1-p)\hat{\rho}\u{f}^\U{B}$ 
and the positivity of $\hat{\rho}\u{f}^\U{B}$, 
we have $p\leq p\u{max}$, 
where $p\u{max}$ is given by Eq.~(\ref{eq:Pmax}) 
with $\alpha'=\beta'=M_1-|m_1|$ and $k=-n=|m_0|+|m_1|$, 
and is strictly positive. 
In such a case, 
the success operator for the gate $\mathcal{M}\u{A}$ 
which achieves the upper bound on the success probability 
is defined by 
\begin{equation}
\hat{M}\u{s}\equiv
{}_{\U{A}}{\braeq{D_{|n|}^{|m_1|}}}{}, 
\label{eq:Ms2}
\end{equation}
which is a linear functional 
but we denote it as a linear operator for convenience. 
From Eqs.~(\ref{eq:dec1}) and (\ref{eq:Ms2}), 
we obtain 
$\hat{M}\u{s}\ket{D_{N}^{M_1}}\u{AB}
=\sqrt{p\u{max}}\ket{D_{N-|n|}^{M_1-|m_1|}}\u{B}$. 
We thus conclude that conditions (\ref{eq:kcondition}) and (\ref{eq:k3})
are sufficient for the transformation, and
the maximum success probability is given by $p\u{max}$ defined in Eq. (\ref{eq:Pmax}).
\section{Universal optimal gates for transforming Dicke states 
by adding one type of spin with minimal access}
A W state is a special case of Dicke states with only one excitation $M_1=1$, {\it i.e.} $\ket{W_N}=\ket{D_N^1}$, recently generated in ion trap~\cite{Iontrap,Iontrap2}, photonic~\cite{WExp,Wex,WExp3} and cavity settings~\cite{WExp2}.
When we expand a W state, a universal optimal gate 
$\mathcal{M}\u{A}$ which achieves the expansion
to $\ket{W_{N+m_0}}=\ket{D_{N+m_0}^1}$~($m_0 > 0$)
by accessing only one qubit is constructed 
as $\hat{M}\u{s}=\ket{W_{n+1}}\u{AA}\bra{1}
+\sqrt{(n+1)^{-1}}\ket{0}^{\otimes n+1}\u{A\ \ \ A}\bra{0}$. 
The expansion can be done
regardless of the size of the W state as 
$\hat{M}\u{s}\ket{W_{N}}=\sqrt{p}\ket{W_{N+n}}$ 
with success probability 
$p=(N+n)N^{-1}(n+1)^{-1}$, 
which coincides with $p\u{max}$ 
calculated from 
Eqs.~(\ref{eq:Pmax}), (\ref{eq:qmin0}) and (\ref{eq:qj}) for any $N$.

Here we show that such a universal optimality is 
partially generalized to Dicke states 
under the following conditions: 
(a) the gate increases at most one type of spin, 
and 
(b) the gate accesses the minimum number of qubits 
to achieve the transformation. 

For a gate with $m_0\leq 0$ and $m_1\leq 0$, 
condition~(b) means that
$k=-n=|m_0|+|m_1|$. 
Then the gate shown in Eq.~(\ref{eq:Ms2}) achieving 
$p\u{max}$ only depends on $m_1$ and $n$. 
Thus it works as a universal optimal gate 
for any input with $M_0\geq |m_0|$ and $M_1\geq |m_1|$,
which is a trivial condition that the
number of spin-up (spin-down) qubits in the input state is at least as big as the number of spin-up (spin-down) qubits to be deleted.

In the case of $m_0>0$, $m_1\leq 0$, and $k=M_1$,
{\it i.e.} adding only spin-down qubits by minimal access,
we have $\alpha'=0$, and $q_j$ defined 
in Eq.~(\ref{eq:qj}) satisfies 
$q_j < q_{j+1}$ 
for $j=0,1,\ldots ,\beta'-1$~\cite{supp}. 
As a result, we have 
$p\u{max}=
q_0C_{N+n}^{k-|m_1|}/C_N^{k}
=
C_{N+n}^{k-|m_1|}/(C_N^{k}C_{k+n}^{k-|m_1|})
$ 
from definition~(\ref{eq:Pmax}). 
We give an explicit construction 
of a universal optimal gate $\mathcal{M}\u{A}^0$ 
which transforms a Dicke state $\ket{D_N^k}$ to $\ket{D_{N+m_0-|m_1|}^{k-|m_1|}}$. 
The gate is characterized by the three parameters $m_0$, $m_1$ and $k$, 
namely $\mathcal{M}\u{A}^0=\mathcal{M}\u{A}^0(m_0, m_1, k)$. 
The gate $\mathcal{M}\u{A}^0$ is a measurement 
represented by 
a success operator $\hat{M}_{\U{s}_0}$ and 
a failure operator 
$\hat{M}_{\U{f}_0} = \sqrt{\hat{I}-\hat{M}_{\U{s}_0}^\dagger\hat{M}_{\U{s}_0}}$, 
and we define $\hat{M}\u{s_0}$ by 
\begin{equation}
\hat{M}\u{s_0}\equiv
\sum_{j=0}^{k-|m_1|}
\sqrt{\frac{C_{k+n}^{k-|m_1|-j}}{C_{k+n}^{k-|m_1|}C_k^{k-j}}}
\keteq{D_{k+n}^{k-|m_1|-j}}\u{AA}\hspace{-0.1cm}\braeq{D_k^{k-j}}. 
\label{eq:Ms0}
\end{equation}
Since $\beta=\min\{ M_0, k\}$ 
and $\beta'=\min\{ M_0, k-|m_1|\}$, 
either $\beta \geq \beta'=k-|m_1|$ or $\beta=\beta'=M_0$ holds. 
Together with $\alpha=\alpha'=0$, 
we see from Eqs.~(\ref{eq:dec1}) and (\ref{eq:Ms0}) that 
\begin{eqnarray}
\hat{M}\u{s_0}\keteq{D_N^k}\u{AB}&=&
\sqrt{\frac{C_{N+n}^{k-|m_1|}}{C_N^{k}C_{k+n}^{k-|m_1|}}}
\keteq{D_{N+n}^{k-|m_1|}}\u{AB}\\
&=&\sqrt{p\u{max}}
\keteq{D_{N+n}^{k-|m_1|}}\u{AB} 
\end{eqnarray}
for any Dicke state $\ket{D_N^k}\u{AB}$. 
Thus the gate $\mathcal{M}\u{A}^0$ 
is the universal optimal gate for transforming Dicke states 
with minimal access of qubits for $m_0 > 0$ and $m_1 \leq 0$. 

In the case of $m_1 > 0$, $m_0\leq 0$ and $k=M_0$,
{\it i.e.} adding only spin-up qubits by minimal access,
we can also construct a universal optimal gate by using the symmetry 
between $\ket{0}$ and $\ket{1}$. 
Let us define a new operator 
$\hat{M}_{\U{s}_1}$ by interchanging the definition of 
$\ket{0}$ and $\ket{1}$ in Eq.~(\ref{eq:Ms0}), 
namely, replacing $m_1$ by $m_0$ and 
$\ket{D_a^b}\u{A}$ by $\ket{D_a^{a-b}}\u{A}$. 
After rewriting the parameter $j$ by $k-j$, we arrive at 
\begin{equation}
\hat{M}\u{s_1}\equiv
\sum_{j=|m_0|}^{k}
\sqrt{\frac{C_{k+n}^{k+m_1-j}}{C_{k+n}^{m_1}C_k^{k-j}}}
\keteq{D_{k+n}^{k+m_1-j}}\u{AA}\hspace{-0.1cm}\braeq{D_k^{k-j}}. 
\label{eq:Ms1}
\end{equation}
By symmetry, the gate 
$\mathcal{M}\u{A}^1$
defined by 
$\hat{M}_{\U{s}_1}$ and 
$\hat{M}_{\U{f}_1} = \sqrt{\hat{I}-\hat{M}_{\U{s}_1}^\dagger\hat{M}_{\U{s}_1}}$ 
achieves the optimal success probability 
$p\u{max}$ when it is applied to $\ket{D_N^{N-k}}\u{AB}$. 

\section{Universal gates for Dicke state transformation}
Here we derive universal gates 
$\mathcal{M}^\U{univ}\u{A}(m_0, m_1, k)\equiv 
\{ \hat{M}^\U{univ}\u{s},\ \hat{M}^\U{univ}\u{f}\}$ 
which transform all Dicke states $\ket{D_N^{M_1}}\u{AB}$ 
satisfying the conditions on $M_0$ and $M_1$
in Eqs.~(\ref{eq:kcondition}) and (\ref{eq:k3}) 
to $\ket{D_{N+n}^{M_1+m_1}}\u{AB}$ 
with nonzero success probabilities. 
For $k=-n=|m_0|+|m_1|$ with $m_0\leq 0$ and $m_1\leq 0$, 
it is easy to see that the gate defined in Eq.~(\ref{eq:Ms2}) 
is a universal gate for any Dicke state 
satisfying $M_0\geq |m_0|$ and $M_1\geq |m_1|$. 
In the following, we therefore consider the case for $k>-n$. 
\begin{figure}[t]
 \begin{center}
  \scalebox{0.45}{\includegraphics{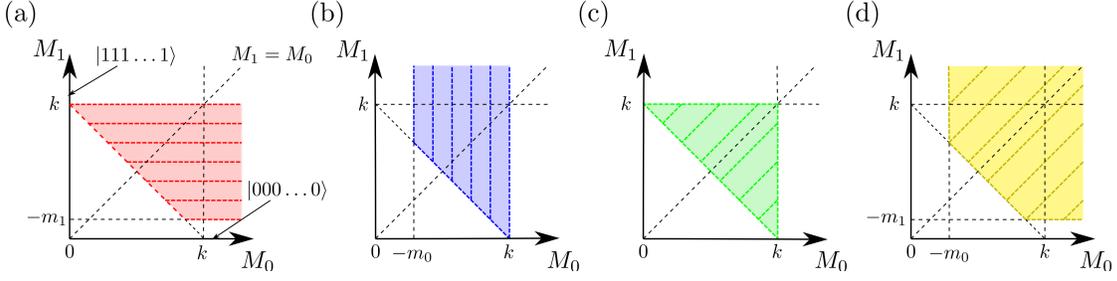}}
 \end{center}
 \caption{
 The successful operating areas of the universal gate 
 $\mathcal{M}\u{A}^\U{univ}$
 defined in Eqs.~(\ref{eq:Ms2}) and (\ref{eq:MsI}) in the case of
 (a) $m_0 > 0$ and $m_1\leq 0$, 
 \textit{i.e.} adding only spin-down qubits,
 (b) $m_0 \leq 0$ and $m_1 > 0$, 
 \textit{i.e.} adding only spin-up qubits,
 (c) $m_0 > 0$ and $m_1 > 0$,
 \textit{i.e.} adding both spin-down and spin-up qubits, and in the case of
 (d) $m_0\leq 0$ and $m_1\leq 0$, 
 \textit{i.e.} only deleting qubits.
 }
 \label{fig:area}
\end{figure}

We define the success operator of the gate by 
\begin{equation}
\hat{M}^\U{univ}\u{s}\equiv
\sum_{j=\alpha\u{s}}^{\beta\u{s}}
\sqrt{q\u{min}^{k}\frac{C_{k+n}^{k+m_1-j}}{C_k^{k-j}}}
\keteq{D_{k+n}^{k+m_1-j}}\u{AA}\hspace{-0.1cm}\braeq{D_k^{k-j}}, 
\label{eq:MsI}
\end{equation}
and define the failure operator by 
$\hat{M}^\U{univ}_{\U{f}} 
= \sqrt{\hat{I}-\hat{M}^\U{univ \dagger}_{\U{s}}\hat{M}^\U{univ}_{\U{s}}}$, 
where 
$\alpha\u{s}\equiv \max\{ 0, -m_0\}$, 
$\beta\u{s}\equiv \min\{ k, k+m_1\}$, and 
\begin{equation}
q\u{min}^{k}\equiv \min_{\alpha\u{s}\leq j\leq \beta\u{s}}
\frac{C_k^{k-j}}{C_{k+n}^{k+m_1-j}} > 0. 
\label{eq:qmin}
\end{equation}
Here the positivity comes from 
$\alpha\u{s}\leq j\leq \beta\u{s}$, implying that 
$0\leq k-j\leq k$ and $0\leq k+m_1-j\leq k+n$. 
In Eq.~(\ref{eq:MsI}), 
by substituting $j=k-M_1+j'$, 
and relabelling $j'$ as $j$, 
$\hat{M}^\U{univ}\u{s}$ is rewritten as 
\begin{equation}
\hat{M}^\U{univ}\u{s}=
\sum_{j=\alpha''}^{\beta''}
\sqrt{q^k\u{min}\frac{C_{k+n}^{M_1+m_1-j}}
{C_k^{M_1-j}}}
\keteq{D_{k+n}^{M_1+m_1-j}}\u{AA}\hspace{-0.1cm}
\braeq{D_k^{M_1-j}}, 
\label{eq:MsI2}
\end{equation}
where 
$\alpha''=\max\{ M_1-k, M_1-k-m_0\}$ and 
$\beta''=\min\{ M_1, M_1+m_1\}$. 
From Eqs.~(\ref{eq:Ms}) and (\ref{eq:MsI2}), 
$\hat{M}^\U{univ}\u{s}$ only differs 
from $\hat{M}\u{s}$ 
by the overall factor $\sqrt{q\u{min}^k/q\u{min}}$ 
and the range of the summation over $j$. 
Because 
$\beta=\min\{ M_1, N-k\}$, 
$\beta'=\min\{ N-k, M_1+m_1\}$ 
and 
$\beta''=\min\{ M_1+m_1, M_1\}$, 
either $\beta'=\beta''$ or $\beta=\min\{ \beta', \beta''\}$ 
is satisfied. 
Similarly, 
because 
$\alpha=\max\{ M_1-k, 0\}$, 
$\alpha'=\max\{ 0, M_1-k-m_0\}$ 
and 
$\alpha''=\max\{ M_1-k-m_0, M_1-k\}$, 
either $\alpha'=\alpha''$ or $\alpha=\max\{ \alpha', \alpha''\}$
is satisfied.  
As a result we have
\begin{eqnarray}
\hat{M}^\U{univ}\u{s}\keteq{D_{N}^{M_1}}\u{AB}
&=&\sqrt{\frac{q\u{min}^k}{q\u{min}}}
\hat{M}\u{s}\keteq{D_{N}^{M_1}}\u{AB}. 
\end{eqnarray}
From 
$\hat{M}\u{s}\ket{D_{N}^{M_1}}\u{AB}=
\sqrt{p\u{max}}\ket{D_{N+n}^{M_1+m_1}}\u{AB}$, 
the success probability of the transformation is 
$p'= p\u{max}q\u{min}^k/q\u{min} $. 
From $q\u{min}^k > 0$, we see that 
the transformation succeeds with a nonzero probability whenever $p\u{max}>0$. 

For convenience, 
we classify the universal gates 
into four cases according to the signs of $m_0$ and $m_1$, 
and show the range of applicable input Dicke states $(M_0, M_1)$
for each case in Fig.~\ref{fig:area}. 
The input states outside of the designated region are not 
transformable by any means~(as $p\u{max}=0$), 
while those in the area are transformed 
with a nonzero success probability by the gate $\mathcal{M}\u{A}^\U{univ}$. 
Thus, taken together the gates we have developed are universal gates for Dicke state transformation. 

\section{Conclusion}
Contrary to the expansion of GHZ, cluster and W states, 
Dicke states cannot be transformed 
by locally accessing only one qubit in general. 
We have derived 
the minimum number of qubits that should 
be accessed to transform a Dicke state to another Dicke state. 
Similarly to the expansion of W states, 
when we access the minimum number of qubits in a physical system for the transformation, 
one can construct universal optimal gates 
which add one type of spin to a given Dicke state. 
We have also constructed 
a universal optimal gate 
which deletes both types of spin from a Dicke state 
with the minimum access of qubits. 
Finally, we have shown the existence of universal gates 
which transform any Dicke state satisfying 
the derived condition for the transformation 
with nonzero probabilities. 
Our results are essential 
for understanding the amount of control  
needed in the preparation and manipulation of 
Dicke states in physical systems such as ion traps, all-optical setups and cavity-QED settings for future quantum information applications.

\section*{Acknowledgement}
This work was supported by the Funding Program 
for World-Leading Innovative R \& D 
on Science and Technology~(FIRST), 
MEXT Grant-in-Aid for Scientific Research 
on Innovative Areas 21102008, 
MEXT Grant-in-Aid for Young scientists(A) 23684035, 
JSPS Grant-in-Aid for Scientific Research(A) 25247068 and (B) 25286077, 
UK EPSRC and the Leverhulme Trust. SKO thanks Dr. Lan Yang for her support.


\begin{thebibliography}{999}
\bibitem{supp} 
From Eq.~(\ref{eq:qj}), we have 
$q_j =
k! (j+m_0)! (k-j-|m_1|)! / (k+n)! j! (k-j)!$.
As a result, from $m_0 > 0$, we have
$q_j / q_{j+1}
=(j+1)(k-j-|m_1|) / (j+1+m_0)(k-j) < 1$, 
namely $q_j < q_{j+1}$ is satisfied.
\end{thebibliography}
\end{document}